\documentclass[traditabstract]{aa}
\usepackage{natbib}
\bibpunct{(}{)}{;}{a}{}{,}
\usepackage{graphics}
\usepackage{epsfig}
\usepackage{txfonts}

\title{The stellar phase density of the local Universe and its
  implications for galaxy evolution}
\titlerunning{The stellar phase density of the local Universe} 
\author{Michael R. Merrifield}
\institute{School of Physics and Astronomy, University of Nottingham,
  University Park, Nottingham, NG7 2RD, UK}

\begin{document}

\abstract{ This paper introduces the idea that the general mixing
  inequality obeyed by evolving stellar phase densities may place
  useful constraints on the possible history of the over-all galaxy
  population.  We construct simple models for the full stellar phase
  space distributions of galaxies' disk and spheroidal components, and
  reproduce the well-known result that the maximum phase density of an
  elliptical galaxy is too high to be produced collisionlessly from a
  disk system, although we also show that the inclusion of a bulge
  component in the disk removes this evolutionary impediment.  In
  order to draw more general conclusions about the evolution of the
  galaxy population, we use the Millennium Galaxy Catalogue to
  construct a model of the entire phase density distribution of stars
  in a representative sample of the local Universe.  In such a
  composite population, we show that the mixing inequality rules out
  some evolutionary paths that are not prohibited by consideration of
  the maximum phase density alone, and thus show that the massive
  ellipticals in this population could not have formed purely from
  collisionless mergers of a low mass galaxy population like that
  found in the local Universe.  Although the violation of the mixing
  inequality is in this case quite minor, and hence avoidable with a
  modest amount of non-collisionless star formation in the merger
  process, it does confirm the potential of this approach.  The future
  measurement of stellar phase densities at higher redshift will allow
  this potential to be fully exploited, offering a new way to look at
  the possible pathways for galaxy evolution, and to learn about the
  environment of star formation through the way that this phase space
  becomes populated over time.  }


\maketitle

\section{Introduction}
\label{sec:introduction}
The stars that make up galaxies exist in a six-dimensional phase space
of positions and velocities, which offers a large amount of freedom in
the possible configurations of stellar populations, and hence the
morphologies and dynamics of the galaxies that they combine to form.
However, because stars do not move instantaneously from one place to
another, and similarly their velocities only change smoothly via
finite gravitational accelerations, the evolution of their
configuration obeys the simple collisionless Boltzmann equation,
\begin{equation}
{d f \over d t} = 0,
\label{eq:CBE}
\end{equation}
which requires that the phase density $f$, the number of stars per
unit volume in both space and velocity, remains constant around any
given star as it travels through space \citep{BinneyTremaine08}.

In principle, this equation places a very strong constraint on the
manner in which a stellar system's properties can evolve with time.
However, there is one significant complication in that although
Eq.~(\ref{eq:CBE}) guarantees that the number of stars within any region
of phase space remains constant, the shape of that region can become
grossly distorted over time, wrapping itself around in a serpentine
fashion.  Just as a confectionery chef lightens the density of toffee
by repeatedly pulling it into strands and wrapping them around, thus
trapping air between the layers of toffee, so the phase wrapping of a
stellar system will create a complex tangle of the original phase
density distribution and empty space.  In practice, one can only hope
to measure the phase space density over a finite region; even in
principle, since stars are intrinsically a discretized sampling of
phase space density, it may not be possible to resolve the complex
phase-wrapped structure. Instead, one measures an average that
combines both the phase density from potentially many original
locations and the mixed-in empty space, creating a ``coarse grained''
distribution function that will always tend to be diluted down from
the original maximum phase density, and thus does not obey
Eq.~(\ref{eq:CBE}).

However, the tendency of this evolution to always dilute the maximum
phase density still places a strong constraint on the possible ways in
which a stellar system can evolve, irrespective of the details of its
evolution.  For example, \citet{Carlberg86} pointed out that the
maximum phase space density in an elliptical galaxy of comparable mass
to the Milky Way is significantly higher than that found in a disk of
similar mass.  It is therefore fundamentally impossible for such a
disk system to be converted into an elliptical through collisionless
processes since there is no way of generating the high phase density
at the centre of the elliptical by mixing the lower phase densities of
a disk.

Although intriguing, this seemingly-fundamental challenge to the
current paradigm in which ellipticals form from mergers of disk systems
has not generally been viewed as a matter of great concern.  As
pointed out by \citet{Lake89}, the controversial region of
high-density phase space represents a tiny fraction of the stellar
distribution right at the extreme maximum of phase density, so that a
relatively minor change to the properties of the initial disk, such as
modestly decreasing its scale-height at large radii, can produce
adequate numbers of stars at high phase density to populate this small
region.  Alternatively, as noted by \citet{Hernquist93} and as we will
also see below, the denser bulge component of a typical disk galaxy
can straightforwardly fill this gap.  Finally, a small amount of star
formation triggered by the transformation process could easily produce
the requisite component at high stellar densities, since the
gas-dynamic processes of star formation are highly collisional and
therefore not subject to this collisionless constraint.  So, there are
a variety of ways to explain away such a violation of the mixing
constraint at this extreme end of the distribution of phase densities,
without any fine tuning of the processes, and thus there is no
compelling reason to throw out the entire paradigm of merger-driven
galaxy evolution on the basis of such an argument.

However, the physics of phase mixing also places limits on possible
evolutionary paths followed by stars that do not lie in this extreme
region of high-density phase space.  The more general criterion was
derived by \citet{Tremaineetal86}, who proved the following theorem.
Define the mass of stars within a galaxy that find themselves in
regions of phase space where the density is greater than $f$ to be
$M(f)$, and the volume in phase space that these stars occupy to be
$V(f)$.  Since both of these quantities vary monotonically with $f$,
one can construct a function $M(V)$ for the system.  For a galaxy to
evolve from an initial form $M_i(V)$ to a final state $M_f(V)$, it is
a necessary condition that
\begin{equation}
M_f(V) \leq M_i(V) \ \ \ \ \forall\ V.
\label{eq:mixing}
\end{equation}
A more recent discussion of this formula and the derivation of an
interesting alternative ``mixing inequality'' can be found in
\citet{Dehnen05}. 

In interpreting the inequality of Eq.~(\ref{eq:mixing}), it is
important to recall that $M$ orders the stars from the highest phase
densities, $f_{\rm max}$, to the lowest, so the smallest values of
$M(V)$ occur at the highest phase densities.  In this region, we can
write $M(V) \approx f_{\rm max} \times V$, so the inequality in
Eq.~(\ref{eq:mixing}) reduces directly to the criterion introduced
above, that the maximum phase density can only decrease.  In
principle, though, we now have a much more general constraint at all
values of $f$.  However, as \citet{Tremaineetal86} noted, the
cumulative nature of $M(V)$ means that, for simple stellar
distributions, if the criterion is met at small values of $V$ (high
phase densities), it will likely be globally met as well.  Thus, in
studying the evolution of simple single model galaxies, such as the
collapse of a uniform spherical distribution to create a credible
elliptical galaxy, \citet{Tremaineetal86} showed that the primary
constraint comes from considering the highest phase density material.

Such simple isolated models of individual galaxies are clearly
somewhat unrealistic.  Real galaxies typically have at least two
stellar components, a disk and a spheroid, which each contribute to
create a rather more complex phase density distribution.  Further, if
we are exploring the role that multiple mergers play in shaping the
morphologies of galaxies, we cannot just consider the properties of
single galaxies in isolation, but rather must look at the phase
density of the entire population.  Fortunately, although the above
formalism was originally considered in the context of the properties
of simple models for individual galaxies, it can equally well be
applied to any self-contained set of stellar systems, including the
entire population of galaxies, each containing multiple components.

In this paper, we begin to develop the methodology for such an
analysis. In Sect.~\ref{sec:model}, we describe the way in which the
phase space density distribution might be reasonably approximated for
the components of each galaxy, and re-derive the original
\citet{Carlberg86} result as a test case.  In
Sect.~\ref{sec:population}, we calculate an initial estimate for the
stellar phase density distribution of the local Universe, and show
that application of the general mixing theorem provides some
interesting pointers on the possible progenitors of elliptical
galaxies.  Section~\ref{sec:discussion} discusses how this methodology
could be taken further.

\section{Stellar phase densities of individual components}
\label{sec:model}
In order to analyze a multi-component picture of the phase space of
galaxies, we need to construct models for the individual components.
Combining these components to derive the global properties of the
population is then reasonably straightforward.  In principle, within a
single galaxy, the populated phase spaces for the different components
can overlap: a bulge star could have a similar position and velocity
to a disk star.  However, the hot nature of the bulge, the rotation of
the disk, and the different spatial scales of the components, mean
that in practice this overlap is small, so, to a reasonable
approximation, $M(f)$ and $V(f)$ can be constructed for a single
galaxy by simple summation of the component bulge and disk parts, and
these quantities can then be further summed to derive the
corresponding functions for the overall population.  Of necessity, the
ingredients of this model will be somewhat simplified, but, as we will
see below, the functions involved stretch over many orders of
magnitude, so uncertainties of factors of a few introduced by any
approximations should not unduly compromise the results.

\subsection{Spheroidal components}
\label{sec:bulgemodel}
The original study by \citet{Carlberg86} modeled elliptical galaxies
as self-gravitating collections of stars following a modified Hubble
profile.  Although this assumption returns a convenient analytical
form for the maximum phase density in such systems, it is somewhat
unrealistic in several aspect.  First, this profile flattens to a
central core, which is not seen in many spheroids: if we are
interested in the regions of highest phase density, it is exactly this
part of the system that we have to consider most closely.  Second, the
assumption that the system is self-gravitating neglects the effects of
any dark matter halo.  This assumption leads to the rather unphysical
corollary that the derived maximum phase density is independent of the
number of stars in the galaxy, so would attribute the same value to
both high and low surface brightness systems.

Here, we adopt a somewhat more observationally-motivated model for
spheroidal components.  We assume that the projected light of a
spheroid can be represented by a de Vaucouleurs law, with effective
radius and luminosity derived from the photometry.  We then
approximate this function using a \citet{Dehnen93} model with
$\gamma=1.5$, thus allowing an analytic formula for stellar density as
a function of radius.  For the velocity distribution, we assume a
Gaussian function with a dispersion $\sigma$ that does not vary with
radius.  Ideally, one would use measured kinematics to determine the
dispersion, but these data do not exist for the whole population, and
such complexity goes beyond the philosophy of this analysis.  Instead,
we use the dispersion predicted by the structural parameters of each
galaxy.  We could use the fundamental plane relations to determine an
appropriate value for $\sigma$, but it was found that in practice this
value was rather sensitive to errors in the photometric decompositions
of galaxies, which occasionally gave implausible values for
scale-lengths and surface brightnesses in bulge components.  However,
the luminosity of the spheroidal component seems to be more robustly
derived in such decompositions, so here we estimate the expected
velocity dispersion from this quantity alone via the Faber--Jackson
relation,
\begin{equation}
\sigma = 10^{-0.1M_B^s + 0.2}\,{\rm km s}^{-1},
\label{eq:FJ}
\end{equation}
where $M_B^s$ is the absolute magnitude of the spheroidal component in the
B-band \citep{ForbesPonman99}.

With the stellar density and velocity distribution determined at each
point in this model for a galaxy's spheroidal component, we can
integrate over the galaxy's phase space to derive the complete
distribution of phase densities, and hence construct $M(f)$ and $V(f)$.

\subsection{Disk components}
\label{sec:diskmodel}
The treatment of disk components is also motivated by the original
analysis of \citet{Carlberg86}, who modeled their spatial distribution
as an exponential in the radial direction, and a self-gravitating
sheet with a constant scale-height in the perpendicular $z$ direction.
The constant scale-height then dictates the radial decrease in the
$z$-component of the velocity dispersion, $\sigma_z$.  If one further
assumes that the other two components of velocity dispersion vary in
the same way with radius, forming a Gaussian velocity ellipsoid, the
phase density distribution across the disk plane is fully specified.
As \citet{Carlberg86} noted, such a distribution has the rather
strange property that the phase density rises with radius in the
galaxy, such that the highest values are found in the outer parts of
the disk.  If this maximum phase density of the disk is to translate
into the maximum phase density of an elliptical galaxy during a
merger, the galaxy must somehow turn itself inside-out.  Although such
a scenario seems unlikely, it is always possible for cold outer parts
of disks to end up forming late in-falling streamers that sink to the
middle of a merger product.  Even the metallicity gradients seen in
ellipticals and disks, which both tend to decrease with radius, do not
present an insurmountable problem given a small amount of star
formation in the merger and a relatively metal-rich bulge in the
original disk galaxy.

We therefore follow this relatively simple prescription of
\citet{Carlberg86}, so that the stellar density is given by
\begin{equation}
\nu(R,z) =
\nu_0 \exp(-R/R_d) {\rm sech}^2(z/z_d),
\end{equation}
with a radial cut-off at $R = 3 R_d$ reflecting that found in most
exponential disks \citep{vanderKruit2001}.  The exact radius and form
for this cut-off turns out not to make much difference to the analysis
since the numbers of stars at such large radii is so small.  The disk
scale-length $R_d$ is obtained from photometric decomposition of the
galaxies we are modeling, while for the scale-height $z_d$ we follow
\citet{MaoMo98} in setting $z_d = 0.2 R_d$ \citep{Bottema93}. For the
kinematic parameters, we adopt the linear fit between velocity
dispersion and B-band disk absolute magnitude, $M_B^d$, from
\citet{Bottema93} to obtain
\begin{equation}
\sigma_z(R=0) = {\rm Max}(-17. \times M_B^d - 280.,10.)\,{\rm km s}^{-1},
\end{equation}
with the limit at small velocity dispersions to ensure that the
central dispersion does not end up smaller than the typical dispersion
of the cold gas from which stars form.  The variation in $\sigma_z$
with radius is then set by the requirement of a constant scale-height,
which implies that it declines exponential with radius with a
scale-length of $2 R_d$.  We further assume, as previous authors have
done, that the shape of the velocity ellipsoid does not vary
significantly with radius, and set $\sigma_z = \sigma_\phi/\sqrt{2} =
\sigma_R/2$, motivated by the values found in the Solar neighbourhood
of the Milky Way, and the epicyclic approximation which fixes the
value of $\sigma_R/\sigma_\phi$ for a flat rotation curve
\citep{BinneyTremaine08}.

Once again, with both photometric and kinematic parameters defined, we
can integrate across phase space to determine the phase density
distribution of any disk component with a specified total luminosity
and radial scale-length, and hence derive $M(f)$ and $V(f)$.

\subsection{A simple application}
\label{sec:MWmodel}
\begin{figure}
 \centerline{\psfig{figure=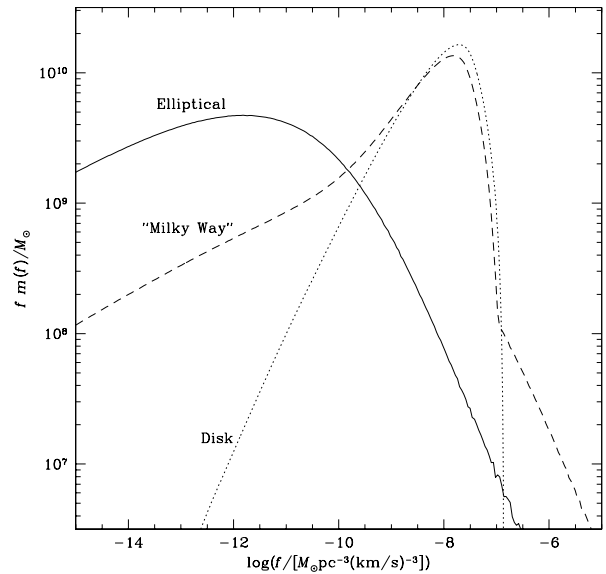,width=8.25cm}} 
  \caption{The distribution of stellar mass as a function of phase
    density, $m(f)$, for model disk and spheroids with a mass of $5 \times
    10^{10} M_\odot$, and for a model of Milky Way-like disk galaxy with both a
    disk and a bulge component.
 \label{fig:mwphase}}  
\end{figure}

As a simple test of these models, we now seek to reproduce the
original result of \citet{Carlberg86} concerning the maximum phase
densities in disks and ellipticals.  Figure~\ref{fig:mwphase} shows
the distribution of phase densities that one obtains for a typical
galactic disk which has been approximately matched to the properties
of the Milky Way, with an absolute magnitude of $M_B^d = -20.0$, a
disk scale-length of $4\,{\rm kpc}$, and a disk mass of $5 \times
10^{10} M_\odot$ \citep{BinneyMerrifield98}.  The figure also shows
the phase density distribution for an identical-mass elliptical galaxy
with an effective radius of $R_e = 10\,{\rm kpc}$.  This phase space
density function has been normalized in the plot such that the area
under the curve is proportional to the mass of stars that find
themselves at a phase density within the range of integration, so
that
\begin{equation}
M(f) = \int_{f}^{\infty} m(f)\,df = \int_{\log f}^{\infty} f m(f) d(\log f).
\end{equation}
As derived by \citet{Carlberg86}, the maximum phase density found in
the elliptical exceeds that of the disk, but, as also noted by
\citet{Lake89}, the area under the curves where this excess occurs is
really very small, so the conflict is fairly insignificant.

\begin{figure}
 \centerline{\psfig{figure=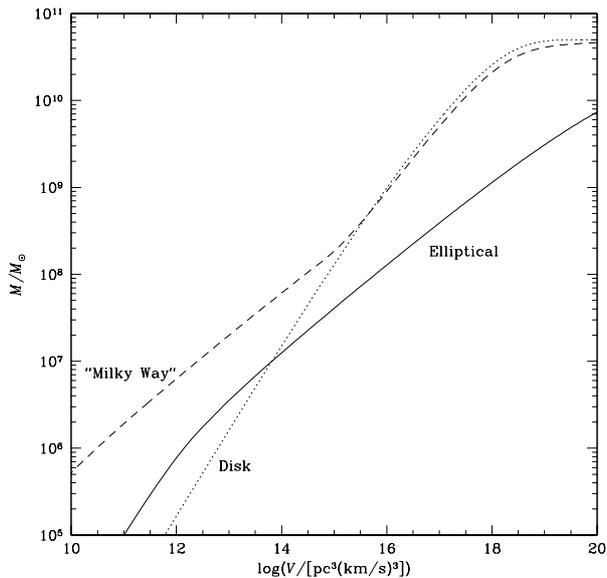,width=8.25cm}} 
 \caption{The cumulative mass in stars below any given phase density
   as a function of the volume of phase space occupied
   by those stars.  The disk, spheroid and Milky Way-like models of
   Fig.~\ref{fig:mwphase} are shown.  To evolve collisionlessly
   from one form to another, a stellar system can only move downward
   in this plot.\label{fig:mwvol}}
\end{figure}

This point is underlined when one considers the more general mixing
constraint on $M(V)$ shown in Fig.~\ref{fig:mwvol}, where it is
clear that rearranging only $\sim 10^7 M_\odot$ of the total $5 \times
10^{10} M_\odot$, or $\sim 0.02\%$ of the stellar mass, at the highest
phase densities (and hence smallest values of $V$) would be sufficient
to eliminate the violation of the mixing inequality.

As Figs.~\ref{fig:mwphase} and \ref{fig:mwvol} further illustrate,
the conflict can also be readily eliminated if one follows the
suggestion of \citet{Hernquist93} and adds to the disk galaxy a bulge
component similar to that found in the Milky Way with a velocity
dispersion of $110\,{\rm km s}^{-1}$, an effective radius of
$2.75\,{\rm kpc}$, and a mass of $1 \times 10^{10} M_\odot$
\citep{BinneyMerrifield98}.  Note that although this model is only
intended to very approximately reproduce the phase-space properties of
the Milky Way's bulge, the maximum phase-space density that it
produces in Fig.~\ref{fig:mwphase} agrees well with the value
of $\sim 10^{-5}M_\odot\,{\rm pc}^{-3}\,({\rm km/s})^{-3}$ inferred by
\citet{Wyse98}.  The maximum phase density decreases with luminosity in
typical spheroidal components \citep{MaoMo98}, so even this relatively
small bulge adds in sufficient stars at high phase densities to
generate the missing high-density extreme of the elliptical.  Thus,
there would be no violation of the mixing constraint of
Eq.~(\ref{eq:mixing}) in turning this disk-dominated galaxy into a
typical elliptical galaxy through collisionless mixing processes.

\section{Stellar phase density of the local Universe}
\label{sec:population}
Although the above test case is interesting, it does not place any
strong limits on the possible general evolutionary paths for galaxies,
since for any individual case one could always find potential
progenitors with structural parameters such that the mixing constraint
is not violated.  It is therefore more interesting to consider the
properties of the entire population of galaxies, to determine more
globally what evolutionary paths the whole population may or may not
have followed.

An important first step in this direction was made by \citet{MaoMo98},
who explored the phase density as a function of absolute magnitude for
a reasonably large sample of disk and elliptical galaxies.  They also
went beyond considering just the extrema of the phase density
distribution by also calculating an average ``effective phase
density'' for these galaxies.  Through this analysis, they were able
to show that disk- and spheroid-dominated galaxies follow distinct
sequences of phase space density as a function of absolute magnitude.
However, they did not possess the decompositions of individual
galaxies into spheroids and disks that would have allowed them to
model multiple components within single systems.  Thus, they were not
in a position to calculate the full stellar phase space density
distribution to ascertain whether, for example, the bulge components
of disk galaxies might be sufficient to eliminate any apparent
violation of mixing constraints, as was illustrated in
Sect.~\ref{sec:MWmodel}.  In addition, the galaxies they analyzed
comprised a somewhat heterogeneous sample, so they did not have access
to a statistically complete sample that would have enabled them to
determine the stellar phase density of the local Universe in a
well-defined volume.

We therefore seek to extend this analysis by considering the data from
the Millennium Galaxy Catalogue\footnote{The Millennium Galaxy
  Catalogue consists of imaging data from the Isaac Newton Telescope
  and spectroscopic data from the Anglo Australian Telescope, the ANU
  2.3m, the ESO New Technology Telescope, the Telescopio Nazionale
  Galileo and the Gemini North Telescope. The survey has been
  supported through grants from the Particle Physics and Astronomy
  Research Council and the Australian Research Council. The data and
  data products are publicly available from
  http://www.eso.org/$\sim$jliske/mgc/.} 
\citep[MGC;][]{Liskeetal03}.  This survey over $37.5\,{\rm deg}^2$,
complete to $m_B = 24$, provides a well-defined and thorough sampling
of galaxies in the local Universe, and the availability of colour data
offers at least a crude conversion from B-band luminosity to stellar
mass \citep{BelldeJong01}, although the connection between these
quantities is necessarily indirect due to the effects of varying
stellar populations and extinction.  Moreover, \citet{Allenetal06}
have shown that the MGC images of these galaxies are of sufficient
quality to be decomposed with some confidence into separate disk and
spheroidal components.  In carrying out such decompositions,
\citet{Driveretal07} found that a relatively modest fraction of
``pseudo-bulges'' are better fitted by a less centrally-concentrated
spheroid function than a de Vaucouleurs law, but that a de Vaucouleurs
law spheroid plus an exponential disk was a generally reasonable
approximation to the observed photometry, allowing us to employ the
simple phase-space components adopted in the current analysis.

Estimating the stellar phase density distribution of the local
Universe from these data is relatively straightforward.  For each of
the 10\,095 galaxies in the MGC for which the decomposition into
exponential disk and de Vaucouleurs spheroid has been made, one
calculates the model phase space densities of the two components as in
Sect.~\ref{sec:model}.  These components are then converted from
luminosity phase densities into mass phase densities using the colour
prescription of \citet{BelldeJong01}.  Finally, the contribution of
each galaxy is weighted according to its total absolute magnitude to
allow for the different volumes sampled by this survey at different
absolute magnitudes, such that each absolute magnitude bin of galaxies
in the MGC contributes a signal proportional to the galaxy luminosity
function at that magnitude, as derived by \citet{Driveretal07}.

\begin{figure}
 \centerline{\psfig{figure=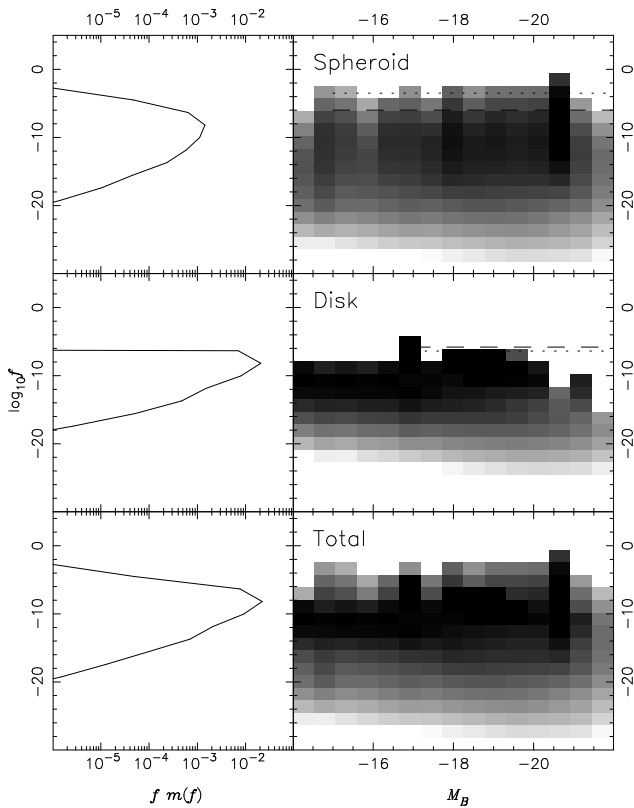,width=8.25cm}} 
 \caption{The phase density of stars in the local Universe as a
   function of the absolute magnitude of the galaxies that produce
   them, as derived from the Millennium Galaxy Catalogue.  The panels
   show both the individual disk and spheroid components and the sum
   of the two.  The projection of these plots by integrating over the
   distribution of galaxy luminosities shows the total stellar phase
   density of the local Universe. Here, $f$ is expressed in $M_\odot
   {\rm pc}^{-3}({\rm km/s})^{-3}$ and $m$ in ${\rm pc}^3 ({\rm km/s})^3
   \,{\rm Mpc}^{-3}$ The dashed and dotted line show, respectively,
   the mean trend in effective and central phase density for
   components considered by \citet{MaoMo98}.\label{fig:MGCphase}}
\end{figure}
The results of this analysis are presented in Fig.~\ref{fig:MGCphase}.
It is notable that, even with the large data set of the MGC, dividing
the sample by luminosity still produces relatively modest numbers of
galaxies per bin, which vary significantly in their properties; these
variations result in the noise apparent in the greyscale image.
However, when the luminosity bins are all combined, this residual
sampling noise is dramatically reduced, producing the reliable smooth
total stellar phase space density distributions shown in the left-hand
panels.

Figure~\ref{fig:MGCphase} also compares these results to the mean
trends established by \citet{MaoMo98}.  For the spheroidal component,
there is good agreement between the two analyses on the upper cut-off
in phase densities.  The effective phase density derived by
\citet{MaoMo98} appears somewhat higher than a measure that one would
infer from the greyscale, but the logarithmic nature of the plot is
somewhat misleading, and it is notable that the value of this
effective phase density does lie very close to the mode in the
projection of the full distribution.

The disk component reveals a similar phenomenon, with the effective
phase density again lying very close to the peak of the distribution.
The central phase density of the disk is less informative, since, as
noted in Sect.~\ref{sec:diskmodel}, it does not represent a maximum
value, but rather the minimum value in the plane of the disk.  Of
course, lower values of phase density can be found away from the plane
of the disk, so this local minimum value has little physical
significance in the context of the total distribution of phase
densities.  The difference between the \citet{MaoMo98} disk relation
and the observed phase density distribution at bright magnitudes is
more interesting, but also has a relatively simple explanation.  In
their analysis, \citet{MaoMo98} selected disk-dominated systems to
derive this relation, so the total absolute magnitude on the abscissa
when plotting their data corresponds quite closely to the disk
luminosity.  However, in the current analysis a random sample of
galaxies from the local Universe has been selected, and at such bright
magnitudes these systems are frequently spheroid-dominated, with
relatively faint but often quite extended disk components.  These
lower-luminosity disks produce the lower phase density of stars that
we see in the disk-component distribution.

The net result of combining these components produces a total stellar
phase density distribution in the local Universe that is quite similar
to that found in a typical spiral galaxy (see Fig.~\ref{fig:mwphase}),
with a degree of structure that can be attributed to the distinct disk
and spheroid components.  This similarity is not so surprising really,
since consideration of the galaxy luminosity function shows that most
of the stellar light comes from galaxies with luminosities around the
break in the Schechter function, and at these luminosities a galaxy is
typically a spiral system like that modeled in Fig.~\ref{fig:mwphase}.

\begin{figure}
 \centerline{\psfig{figure=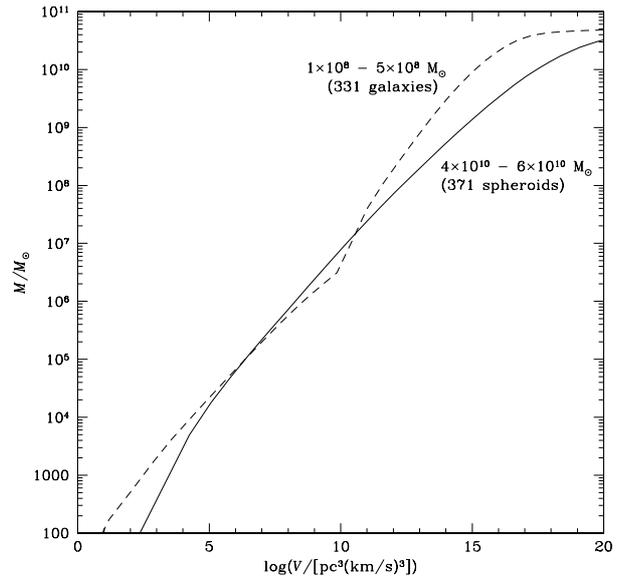,width=8.25cm}} 
 \caption{The cumulative mass function derived from the Millennium
   Galaxy Catalogue for spheroids of mass $\sim 5 \times 10^{10}
   M_\odot$, and for a population of possible progenitor galaxies at
   masses between $1 \times 10^8 M_\odot$ and $5 \times 10^8
   M_\odot$.\label{fig:builde}}
\end{figure}

More interestingly, though, we can now begin to look systematically at
the properties of different sub-classes of galaxies and individual
components.  As a simple illustration, Fig.~\ref{fig:builde} shows
the average cumulative mass function $M(V)$ derived from the phase
space density distribution for the spheroidal components in the local
Universe with masses of $\sim 5 \times 10^{10} M_\odot$.  Also shown
on this figure is the average $M(V)$ curve derived from all the
galaxies in the sample with total stellar masses between $1 \times
10^8 M_\odot$ and $5 \times 10^8 M_\odot$.  The individual galaxy
curves that make up this latter average have been normalized by
multiplying both $M$ and $V$ by the factor required to generate from
them a system containing $\sim 5 \times 10^{10} M_\odot$ of stars.
Thus, it represents the $M(V)$ curve of the progenitors of these
higher-mass spheroids if the latter systems form from random galaxy
mergers of the lower-mass systems.  The fact that $M(V)$ for the
small galaxies does not lie above the curve for the large spheroids at
all values of $V$ means that these functions violate the mixing
inequality of Eq.~(\ref{eq:mixing}), so we can state quite
generally that no combination of collisionless merger processes
bringing small systems like these together could have produced the
final large spheroids.

In this case, it is interesting to note that the issue does not
arise at the highest phase densities (lowest values of $V$), as the
compact spheroidal components of the small galaxies provide more than
enough stars at these densities to produce the bigger spheroids.
Rather it is at intermediate phase densities where one runs out of
spheroid stars, but has yet to tap into the lower phase densities of
disk stars, that the deficit arises.  As such, this apparent violation
of the mixing theorem is somewhat harder to explain away than the
original result comparing a single disk to a single spheroid, where
the discrepancy occurs just in the extreme tail of the phase density
distribution.  

Although this violation of the mixing theorem is formally
statistically significant -- with samples of galaxies of this size,
one can re-sample the distribution to get a good handle on the random
errors in $M(V)$ -- it is nonetheless probably still not a major
problem for the fundamental paradigm of hierarchical galaxy formation.
As with the single galaxy result, it only involves a rather small
fraction of the total mass of the system, and a relatively minor
addition of extra mass through star formation could reorder the
curves.  In addition, presumably today's large spheroids did not form
from progenitors exactly akin to today's small galaxies, so one might
not expect this inequality to be met even if the large spheroids did
form purely from collisionless mergers of pre-existing galaxies.
Indeed, there is now strong evidence that the structural parameters of
even non-star-forming systems have evolved strongly over time
\citep[][and references therein]{Williamsetal10}.  Nonetheless, this
example illustrates the potential power in using phase space
constraints to study the possible evolutionary paths that galaxies
might follow, and where extra stars would have to be added in to allow
other paths to be pursued.  It also underlines the point that there is
more to such analyses than considering just the maximum phase density.

\section{Discussion}
\label{sec:discussion}
In this paper, we have introduced a methodology for modeling the
full stellar phase space density of any combination of disk and spheroidal
stellar systems, including that which makes up the local galaxy
population in the Universe.  We have also discussed the general
inequality that limits the possible ways in which this phase density
distribution can evolve.  As we have seen, such analyses can make use
of more than just the rather non-robust constraint provided by the
maximum phase density, and offer a potentially powerful tool for
determining the possible paths by which these systems could have
evolved.

Of necessity, the phase space model considered here has been rather
simple, but the explosion of available data means that over time many
of the simplifying assumptions can be eliminated.  Large infrared
surveys like Two Micron All-Sky Survey offer a more direct window on
the stellar mass distribution in galaxies, and its decomposition into
disk and bulge components \citep[e.g.][]{MendezAbreuetal08}.  Further,
extensive spectroscopic studies of galaxies using integral field units
\citep[e.g.][]{Krajnovicetal06} and detailed dynamical modeling of
such data \citep[e.g.][]{DeLorenzietal08} mean that the simple
generalized parameterizations of both the photometric and kinematic
properties of galaxies can soon be replaced by reasonably direct
measurements of the phase-space density distribution on a
galaxy-by-galaxy basis.

Another obvious direction in which to extend this analysis is to the
more distant Universe.  At these greater distances, it is challenging
to obtain the necessary high-quality photometric and kinematic
observations, but steps are already being taken both to determine the
distribution of light within such distant galaxies
\citep[e.g.][]{HuertasCompanyetal07}, and in establishing at least the
broad kinematic scaling relations analogous to those used here
\citep[e.g.][]{MacArthuretal08}.  Comparison between the phase density
distributions of stars in distant galaxies and that in the local
Universe will allow a more direct test as to which evolutionary paths
are available to galaxies of different types.  It also offers the
prospect of a new perspective on the star formation history of the
Universe, in that by determining the phase densities of stars that
have to be added over time so as to avoid violation of the mixing
constraint, we will be able to go beyond the simple numbers game of
how many stars form at different epochs to learn about the
smaller-scale environments in which these stars must have formed.

\bibliographystyle{aa}
\bibliography{starphase}

\begin{thebibliography}{23}
\expandafter\ifx\csname natexlab\endcsname\relax\def\natexlab#1{#1}\fi

\bibitem[{{Allen} {et~al.}(2006){Allen}, {Driver}, {Graham}, {Cameron},
  {Liske}, \& {de Propris}}]{Allenetal06}
{Allen}, P.~D., {Driver}, S.~P., {Graham}, A.~W., {et~al.} 2006, \mnras, 371, 2

\bibitem[{{Bell} \& {de Jong}(2001)}]{BelldeJong01}
{Bell}, E.~F. \& {de Jong}, R.~S. 2001, \apj, 550, 212

\bibitem[{{Binney} \& {Merrifield}(1998)}]{BinneyMerrifield98}
{Binney}, J. \& {Merrifield}, M. 1998, {Galactic astronomy} (Galactic astronomy
  / James Binney and Michael Merrifield.~ Princeton, NJ : Princeton University
  Press, 1998.~ (Princeton series in astrophysics) QB857 .B522 1998 (\$35.00))

\bibitem[{{Binney} \& {Tremaine}(2008)}]{BinneyTremaine08}
{Binney}, J. \& {Tremaine}, S. 2008, {Galactic Dynamics: Second Edition}
  (Galactic Dynamics: Second Edition, by James Binney and Scott Tremaine.~ISBN
  978-0-691-13026-2 (HB).~Published by Princeton University Press, Princeton,
  NJ USA, 2008.)

\bibitem[{{Bottema}(1993)}]{Bottema93}
{Bottema}, R. 1993, \aap, 275, 16

\bibitem[{{Carlberg}(1986)}]{Carlberg86}
{Carlberg}, R.~G. 1986, \apj, 310, 593

\bibitem[{{de Lorenzi} {et~al.}(2009){de Lorenzi}, {Gerhard}, {Coccato},
  {Arnaboldi}, {Capaccioli}, {Douglas}, {Freeman}, {Kuijken}, {Merrifield},
  {Napolitano}, {Noordermeer}, {Romanowsky}, \& {Debattista}}]{DeLorenzietal08}
{de Lorenzi}, F., {Gerhard}, O., {Coccato}, L., {et~al.} 2009, \mnras, 395, 76

\bibitem[{{Dehnen}(1993)}]{Dehnen93}
{Dehnen}, W. 1993, \mnras, 265, 250

\bibitem[{{Dehnen}(2005)}]{Dehnen05}
{Dehnen}, W. 2005, \mnras, 360, 892

\bibitem[{{Driver} {et~al.}(2007){Driver}, {Allen}, {Liske}, \&
  {Graham}}]{Driveretal07}
{Driver}, S.~P., {Allen}, P.~D., {Liske}, J., \& {Graham}, A.~W. 2007, \apjl,
  657, L85

\bibitem[{{Forbes} \& {Ponman}(1999)}]{ForbesPonman99}
{Forbes}, D.~A. \& {Ponman}, T.~J. 1999, \mnras, 309, 623

\bibitem[{{Hernquist}(1993)}]{Hernquist93}
{Hernquist}, L. 1993, \apj, 409, 548

\bibitem[{{Huertas-Company} {et~al.}(2007){Huertas-Company}, {Rouan},
  {Soucail}, {Le F{\`e}vre}, {Tasca}, \& {Contini}}]{HuertasCompanyetal07}
{Huertas-Company}, M., {Rouan}, D., {Soucail}, G., {et~al.} 2007, \aap, 468,
  937

\bibitem[{{Krajnovi{\'c}} {et~al.}(2006){Krajnovi{\'c}}, {Cappellari}, {de
  Zeeuw}, \& {Copin}}]{Krajnovicetal06}
{Krajnovi{\'c}}, D., {Cappellari}, M., {de Zeeuw}, P.~T., \& {Copin}, Y. 2006,
  \mnras, 366, 787

\bibitem[{{Lake}(1989)}]{Lake89}
{Lake}, G. 1989, \aj, 97, 1312

\bibitem[{{Liske} {et~al.}(2003){Liske}, {Lemon}, {Driver}, {Cross}, \&
  {Couch}}]{Liskeetal03}
{Liske}, J., {Lemon}, D.~J., {Driver}, S.~P., {Cross}, N.~J.~G., \& {Couch},
  W.~J. 2003, \mnras, 344, 307

\bibitem[{{MacArthur} {et~al.}(2008){MacArthur}, {Ellis}, {Treu}, {U}, {Bundy},
  \& {Moran}}]{MacArthuretal08}
{MacArthur}, L.~A., {Ellis}, R.~S., {Treu}, T., {et~al.} 2008, \apj, 680, 70

\bibitem[{{Mao} \& {Mo}(1998)}]{MaoMo98}
{Mao}, S. \& {Mo}, H.~J. 1998, \mnras, 296, 847

\bibitem[{{M{\'e}ndez-Abreu} {et~al.}(2008){M{\'e}ndez-Abreu}, {Aguerri},
  {Corsini}, \& {Simonneau}}]{MendezAbreuetal08}
{M{\'e}ndez-Abreu}, J., {Aguerri}, J.~A.~L., {Corsini}, E.~M., \& {Simonneau},
  E. 2008, \aap, 487, 555

\bibitem[{{Tremaine} {et~al.}(1986){Tremaine}, {Henon}, \&
  {Lynden-Bell}}]{Tremaineetal86}
{Tremaine}, S., {Henon}, M., \& {Lynden-Bell}, D. 1986, \mnras, 219, 285

\bibitem[{{van der Kruit}(2001)}]{vanderKruit2001}
{van der Kruit}, P.~C. 2001, in Astronomical Society of the Pacific Conference
  Series, Vol. 230, Galaxy Disks and Disk Galaxies, ed. J.~G. {Funes} \& E.~M.
  {Corsini}, 119--126

\bibitem[{{Williams} {et~al.}(2010){Williams}, {Quadri}, {Franx}, {van Dokkum},
  {Toft}, {Kriek}, \& {Labb{\'e}}}]{Williamsetal10}
{Williams}, R.~J., {Quadri}, R.~F., {Franx}, M., {et~al.} 2010, \apj, 713, 738

\bibitem[{{Wyse}(1998)}]{Wyse98}
{Wyse}, R.~F.~G. 1998, \mnras, 293, 429

\end{thebibliography}

\end{document}